



\magnification=\magstep1
\hsize=13cm
\vsize=20cm
\overfullrule 0pt
\baselineskip=13pt plus1pt minus1pt
\lineskip=3.5pt plus1pt minus1pt
\lineskiplimit=3.5pt
\parskip=4pt plus1pt minus4pt

\def\negenspace{\kern-1.1em}



\newcount\secno
\secno=0
\newcount\susecno
\newcount\fmno\def\z{\global\advance\fmno by 1 \the\secno.
                       \the\susecno.\the\fmno}
\def\section#1{\global\advance\secno by 1
                \susecno=0 \fmno=0
                \centerline{\bf \the\secno. #1}\par}
\def\subsection#1{\medbreak\global\advance\susecno by 1
                  \fmno=0
       \noindent{\the\secno.\the\susecno. {\it #1}}\noindent}


\def\sqr#1#2{{\vcenter{\hrule height.#2pt\hbox{\vrule width.#2pt
height#1pt \kern#1pt \vrule width.#2pt}\hrule height.#2pt}}}
\def\square{\mathchoice\sqr64\sqr64\sqr{4.2}3\sqr{3.0}3}


\newcount\refno
\refno=1
\def\y{\the\refno}
\def\myfoot#1{\footnote{$^{(\y)}$}{#1}
                 \advance\refno by 1}


\def\neq{\hbox{$\,$=\kern-6.5pt /$\,$}}





\newcount\secno
\secno=0
\newcount\fmno\def\z{\global\advance\fmno by 1 \the\secno.
                       \the\fmno}
\def\sectio#1{\medbreak\global\advance\secno by 1
                  \fmno=0
       \noindent{\the\secno. {\it #1}}\noindent}


\def\kg{[\![}
\def\gk{]\!]}

\def\la{\langle}
\def\ra{\rangle}




\magnification=\magstep1
\hsize 13cm
\vsize 20cm
\hfill {Preprint IMAFF 95/36}
\bigskip\bigskip
\centerline{\bf{DEGREES OF FREEDOM}}
\centerline{\bf{OF}}
\centerline{\bf{ARBITRARILY HIGHER-DERIVATIVE FIELD THEORIES}}
\vskip 0.3cm
\centerline{by}
\vskip 0.7cm
\centerline{F.J. de Urries}
\vskip 0.2cm
\centerline{\it Departamento de F\'\i sica, Universidad de Alcal\'a de
            Henares,}
\centerline{\it 28871 Alcal\'a de Henares (Madrid), Spain}
\centerline{and}
\centerline{\it{IMAFF, Consejo Superior de Investigaciones
Cient\'\i ficas,}}
\centerline{\it{Serrano 123, Madrid 28006, Spain}}
\vskip 0.7cm
\centerline{J.Julve}
\vskip 0.2cm
\centerline{\it{IMAFF, Consejo Superior de Investigaciones
Cient\'\i ficas,}}
\centerline{\it{Serrano 123, Madrid 28006, Spain}}

\vskip 0.7cm

\centerline{ABSTRACT}\bigskip
As an example of what happens with physically relevant theories
like effective gravity, we consider the covariant relativistic
theory of a scalar field of arbitrarily higher differential
order. A procedure based on the Legendre transformation and
suitable field redefinitions allows to recast it as a theory of
second order with one explicit independent field for each degree
of freedom. The physical and ghost fields are then apparent.
The full (classical) equivalence of both
Higher and Lower Derivative versions is shown. An artifact of
the method is the appearance of irrelevant spurious fields which
are devoid of any dynamical content.

\vfill
\eject

\bigskip\bigskip
\sectio{\bf{Introduction}}\bigskip
Theories of gravity with terms of any order in curvatures
arise as part of the low energy effective theories of the
strings [1] and from the dynamics of quantum fields in a
curved spacetime background [2].

Theories of second order (4--derivative theories in the
following) have been studied more closely in the literature
because they are renormalizable [3] in four dimensions and have nice
renormalization group properties [4]. In particular a procedure
based on the Legendre transformation was devised [5] to recast
them as an equivalent theory of second differential order.
A suitable diagonalization of the resulting theory was found
later [6] that yields the explicit independent fields for the
dynamical degrees of freedom involved. In [7] the simplest
example of this procedure was given using a model of one
scalar field with a massless and a massive degree of freedom.
In an appendix, Barth and Christensen [8] gave the splitting of the
higher derivative (HD)
propagator into quadratic ones for the 4th, 6th and 8th
differential-order scalar theories though not devising the
systematic procedure for the general case.

The problem remained open of how to tackle arbitrary
derivative-order theories, as it is the general case, since the
Legendre transformation procedure then becomes far from trivial.
Classical treatises [9] face the Lagrangian and Hamiltonian
theories of systems including higher time derivatives
of the generalized coordinates and the definition of canonical
momenta. Later work has considered the variational problem of
those theories with the tools of the Cartan form, k-jets, symplectic
geometry and Legendre mappings [10]. The difficulties of the
seemingly unavoidable trading of unitarity against non locality
have been also studied [11]. Recently, S.Hamamoto [12]
has proven the equivalence of the path integral for the theory
formulated in terms of constrained systems and Dirac's method,
and for the classical Ostrogradski's treatment. However the
particular case of relativistic covariant
field theories, though involving only even differential orders, has
complications of its own and, of course, is not trivially
covered by those general treatments.

We address this issue by using a simplified model with
scalar fields as in [7][8]. Our presentation highlights the Lorentz
covariance and the particle aspect of the
theory, with emphasis in the structure of the propagators and the
coupling to other matter sources. In Section 2 we study the case of the
4-derivative theory for arbitrary masses, which exemplifies the
use of the Helmholtz Lagrangian and the crucial diagonalization
of the fields. In Section 3 we work out the 6-derivative case
where the complications characterising the general case appear
for the first time, including the occurrence of a new family of
spurious fields. The notation needed to deal with the previous
case is fully generalized in Section 4 where the general
2N-derivative theory is considered. Four Appendices are devoted
to the technical details of some proofs and the cohomological
interpretation of a result.

\vfill
\eject

\sectio{\bf Four-derivative theory and notations.}
\bigskip

In this section we will introduce a convenient notation, and try
to get the reader acquainted with our treatment.

We adopt the Minkowski signature $(1,-1,-1,-1)$.

Masses are ordered such that $ m_i > m_j $ when $ i < j$.

$\kg i\gk \equiv (\square + m^2_i)$, the Klein--Gordon (KG) operator for mass
$ m_i $ .

$ \langle ij \rangle \equiv (m^2_i - m^2_j)$ is positive when $i < j$. It will
always be written with this

ordering unless we wish to highlight some symmetry property.

Notice that
$ \kg j\gk = \kg i\gk - \langle ij \rangle$.
\bigskip\bigskip

Here we generalize the example in [7] to arbitrary masses.
Consider the 4--derivative scalar theory

$$ {\cal L}^{(4)} = -{1 \over 2}{1 \over {(m_1^2 - m_2^2)}}
              \,\phi^o(\square + m_1^2)(\square + m_2^2) \phi^o
                                          -j\,\phi^o\quad.\eqno(\z)$$

\noindent{It} yields the propagator

$$-{{(m_1^2 - m_2^2)}\over{(\square + m_1^2)(\square + m_2^2)}}
      = {1 \over{\square + m_1^2}} - {1 \over{\square + m_2^2}}
                                        \quad, \eqno(\z)$$

\noindent{where} the pole at $m_2$ is physical and the one at
$m_1$ is a poltergeist.

Dropping total derivatives Eq.(2.1) may be written as

$$ {\cal L}^{(4)}[\phi^o,\kg 1\gk\phi^o]
                     =-{1 \over 2}{1 \over{\langle 12 \rangle}}
                                      \bigl[(\kg 1\gk\phi^o)^2
             - \langle 12 \rangle \phi^o(\kg 1\gk\phi^o)\bigl]
                                    -j\,\phi^o\quad .\eqno(\z)$$

\noindent{Define} the canonical conjugate variable

$$ \pi \equiv
          {{\partial{\cal L}^{(4)}}\over{\partial([\![1]\!]\phi^o)}}
            =-{1\over{\langle 12 \rangle}}\bigl[[\![1]\!]\phi^o
                      - {1\over 2}\langle 12 \rangle\phi^o\bigl]
                                              \quad ,\eqno(\z)$$

\noindent{from} which

$$ [\![1]\!]\phi^o = \langle 12 \rangle(-\pi
                               + {1\over 2}\phi^o)\equiv {\bf Q}
                                             \quad . \eqno(\z)$$

\noindent{The} "Hamiltonian" is

$$ {\cal H}[\phi^o,\pi] \equiv \pi {\bf Q} -
                                  {\cal L}^{(4)}[\phi^o,{\bf Q}]
                                               \quad ,\eqno(\z)$$

\noindent{that} is

$$ {\cal H} =-{1\over 2}\langle 12 \rangle
                                     (-\pi + {1\over 2}\phi^o)^2
                                      +j\,\phi^o\quad .\eqno(\z)$$

\noindent{The} Helmholtz Lagrangian (the Euler equations of which
are the canonical ones from $\cal H$) is

$${\cal L}^H \equiv \pi [\![1]\!]\phi^o - {\cal H}\quad .\eqno(\z)$$

\noindent{Upon} diagonalization by changing to new variables

$$\eqalign     {\phi^o &= \phi^1_1 + \phi^1_2 \cr
                 \pi &= {1\over 2}(\phi^1_1 - \phi^1_2)\quad, \cr}
                                              \eqno(\z)$$

\noindent{or} conversely

$$\eqalign      {\phi^1_1 &= \pi + {1\over 2}\phi^o \cr
                 \phi^1_2 &= -\pi + {1\over 2}\phi^o\quad, \cr}
                                                 \eqno(\z)$$

\noindent{one} obtains the equivalent 2--derivative theory

$${\cal L}^{(2)} = {1\over 2}\,\phi^1_1 [\![1]\!]\phi^1_1
                  -{1\over 2}\,\phi^1_2 [\![2]\!]\phi^1_2
                                 -j\,(\phi^1_1 + \phi^1_2)
                                           \quad .\eqno(\z)$$

\noindent{The} propagators stemming from (2.11) are just the
ones in the r.h.s. of (2.2). Notice that our notation is such
that, here and in the following, the physical (negative) sign is
beared by the lightest field, namely  $\,\phi^1_2\,$ or generally the
field with the highest subindex. This seemingly unnecessary
wealth of upper and lower index labels has been introduced for
further generalization.

The splitting of the quartic propagator displayed in (2.2)
tells us that the emission of a "particle" endowed with such a
propagator is actually equivalent to the emission of two
particles (one physical and one with nonphysical norm) with
quadratic propagators. They are made explicit in (2.11) by two
independent fields the sum of which couples to the source.

\bigskip

The equivalence between the equations of motion can be tested as
well. For (2.1) the HD Euler equation is

$$ -{1 \over \langle 12 \rangle}\,[\![1]\!][\![2]\!]\phi^o = j
                                             \quad . \eqno(\z)$$

\noindent{Hamilton's} canonical equations from (2.6) are, by definition,
the same as Euler's from (2.8) for $\phi^o$ and $\pi$, which
respectively are

$$\eqalign{[\![1]\!]\pi &= -{1 \over 2}\langle 12 \rangle
                           (-\pi + {1\over 2}\phi^o) + j \cr
        [\![1]\!]\phi^o &= \langle 12 \rangle
                           (-\pi + {1\over 2}\phi^o)\quad.\cr}
                                               \eqno(\z)$$

\noindent{Finally} for the lower derivative (LD) theory
(2.11) one has

$$\eqalign{  [\![1]\!]\phi^1_1 &= j \cr
            -[\![2]\!]\phi^1_2 &= j\quad. \cr}\eqno(\z)$$

\noindent{By} changing variables according to (2.10) one may
recast (2.14) into (2.13). Then working $\pi$ out of the 2nd
equation
(2.13) and substituting in the 1st one, after a little algebra one
recovers (2.12). This proves the full (classical) equivalence of
the theories (2.1) and (2.11).
\bigskip

\bigskip\bigskip

\sectio{\bf 6-derivative theories}
\bigskip

The conveniently normalized 6-derivative Lagrangian is

$$ {\cal L}^6_{\phi^o} = - {1\over 2}
{1\over {\langle 12\rangle\langle 13 \rangle\langle 23\rangle}}
                \,\phi^o\kg 1\gk\kg 2\gk\kg 3\gk\phi^o -j\phi^o
                                               \quad , \eqno(\z)$$

\noindent{which} may be rewritten as

$$  {\cal L}_{\psi}  = - {1\over 2}
{1\over {\la 12\ra\la 13\ra\la 13\ra}}\,\psi\kg 1\gk\kg 2\gk\psi -
                                     j{\kg 3\gk}^{-{1\over 2}}\psi
                                                    \quad , \eqno(\z)$$

\noindent{where} $\psi\equiv{\kg 3\gk}^{1\over 2}\phi^o$.
The Lagrangian (3.1) has mass dimension 2, so a further
dimensional constant in front of it must be understood.

Applying the procedure described in the previous section to (3.2),
with the diagonalization

$$\eqalign { \psi^1_1 &= \pi + {1\over 2}\psi \cr
             \psi^1_2 &= -\pi + {1\over 2}\psi \quad,\cr}
                                                  \eqno(\z)$$
\noindent{one} obtains

$${\cal L}_{\psi^1_1\psi^1_2} = {1\over 2}{1\over {\la
                       13\ra\la 23\ra}}\,\psi^1_1\kg 1\gk\psi^1_1 -
                       {1\over 2}{1\over {\la
                       13\ra\la 23\ra}}\,\psi^1_2\kg 2\gk\psi^1_2 -
                      j{\kg 3\gk}^{-{1\over 2}}(\psi^1_1 + \psi^1_2)
                                                 \quad , \eqno(\z)$$
\noindent{that} with

$$\eqalign{ \phi^1_1 &= {\kg 3\gk}^{-{1\over 2}}\psi^1_1 \cr
            \phi^1_2 &= {\kg 3\gk}^{-{1\over 2}}\psi^1_2
                                       \quad,\cr}\eqno(\z)$$
\noindent{gives}

$${\cal L}^4_{\phi^1_1\phi^1_2} =
                          {1\over 2}{1\over {\la 13\ra\la 23\ra}}\,
                                   \phi^1_1\kg 1\gk\kg 3\gk\phi^1_1
                        - {1\over 2}{1\over {\la 13\ra\la 23\ra}}\,
                                   \phi^1_2\kg 2\gk\kg 3\gk\phi^1_2
                                           - j(\phi^1_1 + \phi^1_2)
                                                \quad . \eqno(\z)$$

We now repeat the procedure of Section 2 for both fields $\phi^1_1$ and
$\phi^1_2$. Notice that the factors ${1\over {\la
23\ra}}$ and ${1\over {\la 13\ra}}$  are spectators when
performing this process for the first and second fields
respectively. The diagonalizations to be made now are given by

$$\eqalign{\phi^2_1 &= \pi^1_1 + {1\over 2} \phi^1_1\cr
           \Phi^2_3 &= -\pi^1_1 + {1\over 2}\phi^1_1 \quad,\cr}
                                                  \eqno(\z)$$

\noindent{and}

$$\eqalign{\phi^2_2 &= \pi^1_2 + {1\over 2} \phi^1_2\cr
           \Theta^2_3 &= -\pi^1_2 + {1\over 2}\phi^1_2 \quad,\cr}
                                                   \eqno(\z)$$

\noindent{that} lead to the following 2-derivative Lagrangian

$$\eqalign{{\cal L}^2 =
               -{1\over 2}{1\over \la 23\ra}\phi^2_1\kg 1\gk\phi^2_1
               +{1\over 2}{1\over \la 23\ra}\Phi^2_3\kg 3\gk\Phi^2_3
              &+{1\over 2}{1\over \la 13\ra}\phi^2_2\kg 2\gk\phi^2_2
         -{1\over 2}{1\over \la 13\ra}\Theta^2_3\kg 3\gk\Theta^2_3\cr
       &-j(\phi^2_1 + \phi^2_2 + \Phi^2_3 +\Theta^2_3)\quad.\cr}
                                                        \eqno(\z)$$

The surprise has come up of the duplication of particles with
the same mass $m_3$. This can be dealt with by observing that
only the linear combination $\phi^2_3 = \Phi^2_3 + \Theta^2_3$
couples to the source, whereas the linearly independent one
$\zeta^2_3 = C_2 {{\la 13\ra}\over {\la 23\ra}}\Phi^2_3 + C_2 \Theta^2_3$
is decoupled, where $C_2$ is real and $\not=0$. In terms of
these new fields  the theory gets its final transparent form

$$\eqalign{{\cal L}^2 =
               -{1\over 2}{1\over \la 23\ra}\phi^2_1\kg 1\gk\phi^2_1
               +{1\over 2}{1\over \la 13\ra}\phi^2_2\kg 2\gk\phi^2_2
              &-{1\over 2}{1\over \la 12\ra}\phi^2_3\kg 3\gk\phi^2_3
                                -j(\phi^2_1 + \phi^2_2 + \phi^2_3)\cr
          &+{1\over 2}{{\la 23\ra}\over{(C_2)^2\la 12\ra\la 13\ra}}
                              \zeta^2_3\kg 3\gk\zeta^2_3\quad.\cr}
                                                         \eqno(\z)$$

Contrarily to the expectations we have ended up with four
degrees of freedom instead of three. However the field
$\zeta^2_3$ is devoid of any dynamical content since it does not
couple either to the source or to the other fields, it may then
be arbitrarily normalized and does not propagate between
sources. Therefore it must be regarded as an spurious field. A
trivial way to dispose of it is realizing that (3.10) is invariant
under the local Abelian transformations
$\delta\zeta^2_3 = \lambda$, with $\lambda$ obeying
$\kg 3\gk \lambda(x) = 0$, and using them to gauge away the
field.
\bigskip

The equations of motion of the 2-derivative theory (3.10) are

$$\eqalign{ -{1\over{\la 23\ra}}\kg 1\gk\phi^2_1& = j \cr
             {1\over{\la 13\ra}}\kg 2\gk\phi^2_2& = j \cr
            -{1\over{\la 12\ra}}\kg 3\gk\phi^2_3& = j\quad, \cr}
                                                \eqno(\z)$$

\noindent{plus} the trivial decoupled one

$$\kg 3\gk\zeta^2_3 = 0 \quad . \eqno(\z)$$

\noindent{Recovering} the HD equation of motion corresponding to
(3.1), namely

$$ -{1\over{\la 12\ra\la 13\ra\la 23\ra}}\,
                       \kg 1\gk\kg 2\gk\kg 3\gk\phi^o = j \quad ,
                                                       \eqno(\z)$$

\noindent{can} be achieved by adding and substracting the LD
ones (3.11) and (3.12) between themselves and undoing the
various diagonalizations and field redefinitions done at several
stages.

Here also the dynamically irrelevant role of $\zeta^2_3$ is
shown by the fact that its equation of motion adds or substracts
0 to the others in the first step.
Moreover, even a coupling $-\lambda j\zeta^2_3$ added to (3.10)
would be
immaterial as long as the ensuing modification of the equation
of motion (3.12) cancels out when going back to (3.13). This is
shown in Appendix A.
\bigskip

Finally the propagators stemming from (3.10) are the pieces
found in the algebraic splitting of the HD one, namely

$$-{{\la 12\ra\la 13\ra\la 23\ra}
    \over {\kg 1\gk\kg 2\gk\kg 3\gk}} =
                               -{{\la 23\ra}\over{\kg 1\gk}}
                               +{{\la 13\ra}\over{\kg 2\gk}}
                               -{{\la 12\ra}\over{\kg 3\gk}}
                                           \quad .\eqno(\z)$$

\noindent{This} completes the proof of the full (classical)
equivalence of both theories (3.1) and (3.10).

\bigskip
\sectio{\bf{2N-derivative general theory.}}

Once the 6-derivative theory has been worked out, we face
the general 2N-derivative case along the same lines.
The Lagrangian is

$$ {\cal L}^{2N}_{\phi^o} = - {1\over 2}
{1\over {\prod\limits_{(ij)=(12)\atop{ }}^{{ }\atop(N-1\,N)}\;
          \la ij\ra}}\;
            \phi^o\prod\limits_{i=1}^{N}\kg i\gk\phi^o -j\phi^o
                                           \quad . \eqno(\z)$$

\noindent{The} product in the denominator must be calculated for
all the ordered pairs
$(ij)$ ranging from $(12)$ to $(N\!-\!1\;N)$ with $i<j$ , so that
always $\la ij\ra>0$.  A dimensional constant has been omitted.

The HD propagator stemming from (4.1) can be expanded as
follows

$$-\,{{\prod\limits_{(ij)=(12)\atop{ }}^{{ }\atop(N-1\,N)}\;
          \la ij\ra} \over{\prod\limits_{i=1}^{N}\kg i\gk}} =
                              \sum_{i=1}^{N}\;(-1)^{N+i+1}
                {{\prod\limits_{ (kl)=(12)\atop k,l\not=i}^
                     {{ }\atop(N-1\,N)}\;\la kl\ra} \over
                          {\buildrel { }\over{\kg i\gk}}}
                                      \quad . \eqno(\z)$$

On the r.h.s. of (4.2), alternating plus and minus signs occur,
the one for the
smallest mass term being negative. So (4.2) gives us the
splitting of the propagator with N poles in terms of simple
Klein-Gordon quadratic propagators. The signs give unphysical caracter to
many poles, but it is always possible to choose the one with the
smallest mass as a physical pole, as done here.

In order to prove (4.2), we follow the induction method.
Assume it holds for N, then for N+1 we have

$$\eqalign{
-\,{{\prod\limits_{(ij)=(12)}^{{ }\atop(N\,N+1)}\;\la ij\ra}\over
  {\prod\limits_{i=1}^{{ }\atop N+1}\kg i\gk}}
&=
-\,{{\prod\limits_{(kl)=(12)}^{{ }\atop(N-1\;N)}\;\la kl\ra}\over
  {\prod\limits_{i=1}^{N}\kg i\gk}}\quad
          {{\prod\limits_{m=1\atop{ }}^{N}\;\la m\;N+1\ra}\over
               {\buildrel { }\over{\kg N+1\gk}}}\cr
&= \sum_{i=1}^{N}\;(-1)^{N+i+1}
   {{\prod\limits_{ (kl)=(12)\atop k,l\not=i}^
   {{ }\atop (N-1\;N)}\;\la kl\ra} \over{\kg i\gk}}
          {{\prod\limits_{m=1\atop{ }}^{N}\;\la m\;N+1\ra}\over
                         {\buildrel{ }\over{\kg N+1\gk}}}\cr
&= \sum_{i=1}^{N}\;(-1)^{N+i+1}
  {\prod\limits_{ (kl)=(12)\atop k,l\not=i}^{{ }\atop (N-1\;N)}\; \la kl\ra}
   {\prod\limits_{m=1}^{N}\;\la m\;N+1\ra}
           \Biggl({1\over{\buildrel{ }\over{\kg i\gk\kg
                                  N+1\gk}}}\Biggl)\quad,\cr}
                                                   \eqno(\z)$$

\noindent{where} we have used (4.2) for the second step.

Taking into account that $i\leq N$ , and that as done in (2.2)
for the poles $1$ and $2$, for $i<j$ one has that

$$ {1\over{\kg i\gk\kg j\gk}} = - {1\over{\la ij \ra}}
                                             {1\over{\kg i\gk}}
                                 + {1\over{\la ij \ra}}
                                             {1\over{\kg j\gk}}\quad ,
                                                    \eqno(\z)$$

\noindent{the} last expression in (4.3), may be written as

$$ \eqalign{
  & \sum_{i=1}^{N}\;(-1)^{N+1+i+1}
  {\prod\limits_{ (kl)=(12)\atop k,l\not=i}^{{ }\atop (N-1\;N)}\; \la kl\ra}
   {\prod\limits_{m=1\atop m\not=i}^{N}\;\la m\;N+1\ra}
        {1\over{\buildrel{ }\over{\kg i\gk}}}\cr
+ \Biggl( &\sum_{i=1}^{N}\;(-1)^{N+1+i}
  {\prod\limits_{ (kl)=(12)\atop k,l\not=i}^{{ }\atop (N-1\;N)}\; \la kl\ra}
   {\prod\limits_{m=1\atop m\not=i}^{N}\;\la m\;N+1\ra} \Biggl)
   {1\over{\buildrel{ }\over{\kg N+1\gk}}}\quad.\cr}
                                             \eqno(\z)$$

Now, in both terms the product symbols can be merged into one
symbol so that (4.5) becomes

$$
   \sum_{i=1}^{N}\;(-1)^{N+1+i+1}
 {{\prod\limits_{(kl)=(12)\atop k,l\not=i}^{{ }\atop (N\;N+1)}\; \la kl\ra}
  \over{\buildrel{ }\over{\kg i\gk}}}
+ \Biggl( \sum_{i=1}^{N}\;(-1)^{N+1+i}
  {\prod\limits_{(kl)=(12)\atop k,l\not=i}^{{ }\atop (N\;N+1)}\; \la kl\ra}
   \Biggl)
   {1\over{\buildrel{ }\over{\kg N+1\gk}}}  \quad .
                                             \eqno(\z)$$

\noindent{Next} the 2nd term in (4.6) could be embodied in the 1st
one by extending there the summation to the value $N+1$, (4.6) then
becoming

$$  \sum_{i=1}^{N+1}\;(-1)^{N+1+i+1}
 {{\prod\limits_{(kl)=(12)\atop k,l\not=i}^{{ }\atop (N\;N+1)}\; \la kl\ra}
  \over{\buildrel{ }\over{\kg i\gk}}}  \quad,
                                        \eqno(\z)$$

\noindent{provided} we could show that

$$ \sum_{i=1}^{N}\;(-1)^{N+1+i}
  {\prod\limits_{(kl)=(12)\atop k,l\not=i}^{{ }\atop (N\;N+1)}\; \la kl\ra}
= (-1)^{N+N+3}{\prod\limits_{(kl)=(12)}^{{ }\atop (N-1\;N)}\;
\la kl\ra}  \quad .       \eqno(\z)$$

But (4.7) equals the l.h.s. of (4.3), so (4.2) is proved for $N+1$
once (4.8) has been shown to hold. This is done in Appendix B.
{}From (4.8), an interesting cohomological result is obtained in
Appendix C.

\bigskip

Once the validity of the splitting formula for the 2N-order
propagator has been shown, we steer to the problem of deriving
the LD Lagrangian that yields the quadratic propagators.

The starting HD lagrangian (4.1), namely

$${\cal L}^{2N}=
        -{1\over 2}{\prod\limits_{(ij)=(12)}^{{ }\atop (N-1\;N)}}
                     {1\over{\la ij\ra}}\;\phi^o\kg 1\gk\kg
              2\gk\cdot\cdot\cdot\kg N\gk\phi^o\; - j\,\phi^o
                                         \quad ,\eqno(\z)$$

\noindent{can} be handled, as in the case
of the 6-derivative theory, by successive Legendre
transformations and one ends up with the following 2-derivative theory:

$$\eqalign{
    {\cal L}^2 =&{1\over 2}\sum_{i=1}^{{ }\atop N}(-1)^{N-i+1}
    \Biggl({\prod\limits_{(mn)=(12)\atop m,n\not= i}^{{ }\atop (N-1\,N)}}
                              {1\over{\la mn\ra}} \Biggl)
                          \phi_i^{N-1}\kg i\gk\phi_i^{N-1}
   -j\,\Biggl( \sum_{i=1}^{{ }\atop N}\phi_i^{N-1}\Biggl)\cr
                       + &{1\over 2}\sum_{M=3}^{{ }\atop N}
                 \Biggl(\sum_{l=1}^{(2^{M-2}-1)}(-1)^{N-M+l-1}
        \zeta_{Ml}^{N-1}\kg M\gk\zeta_{Ml}^{N-1}\Biggl)\quad.\cr}
                                                   \eqno(\z)$$

Here, the upper and lower indices in the fields $\phi_i^{N-1}$
stand to indicate that they are obtained from $\phi^o$  after
$N-1$ Legendre transformations and have mass $m_i$. These fields
couple to the source, and their free Lagrangians exactly fit what
is needed to get the particle poles occurring in the r.h.s. of
(4.2). Therefore the degrees of freedom are conserved and the
physical or ghostly character of the fields in (4.10) are the same
as in (4.2).
A crowd of spurious fields $\zeta_{Ml}^{N-1}$ arise, but again they are
irrelevant. They are degenerate in mass, being their number
$(2^{M-2}-1)$ for the mass $m_M$, where $M$ ranges from 3 to
$N$, and their Lagrangians bear the
sign given in (4.10). The proof of the dynamical equivalence of
(4.10) and (4.9) is carried out in Appendix D.
\bigskip

The LD equations of motion, namely

$$\eqalign{
 (-1)^{N-i+1}\Biggl({\prod\limits_{(mn)=(12)\atop m,n\not= i}
                                 ^{{ }\atop (N-1\,N)}}
       {1\over{\la mn\ra}} \Biggl)\kg i\gk\phi_i^{N-1}\,&=\,j\quad
                                                    (i=1,...,N)\cr
 (-1)^{N-M+l-1}\kg M\gk\zeta_{Ml}^{N-1}\,&=\,0\quad
               {{M=3,...,N}\choose{l=1,...,2^{M-1}-1}}\quad,\cr}
                                                    \eqno(\z)$$

\noindent{can} be traced back to the HD one

$$   -{\prod\limits_{(ij)=(12)}^{{ }\atop (N-1\;N)}}
                     {1\over{\la ij\ra}}\;\kg 1\gk\kg
              2\gk\cdot\cdot\cdot\kg N\gk\phi^o\,=\,j\quad,
                                                   \eqno(\z)$$

\noindent{as} in the 4-derivative and 6-derivative cases. This
establishes the full classical equivalence between the HD and the
LD theories also for the general 2N-derivative case.

\bigskip\bigskip

\sectio{\bf{Conclusions}}\bigskip

Starting from a general HD relativistic covarianttheory for a
scalar field, we have devised the procedure for translating it into an
equivalent 2-derivative theory with as many independent scalar
fields as degrees of freedom the HD theory had. By studying the
equations of motion we have assessed the full classical
equivalence of both versions of the theory. The physical picture
stemming from this result is that the emission of one "particle" by
a source in a 2N-derivative theory, is equivalent to the emission of
N particles described by the usual Klein-Gordon 2-derivative theory.

The procedure followed here, based on the Legendre transformation,
works only when all the masses involved are different,
many expressions becoming singular otherwise as a cosequence of
the system not being regular. The case of the (conformally
invariant in four dimensions) HD theories of gravity based on
the squared Weyl tensor, where only the highest derivative terms
occur since all the masses are zero, has this kind of difficulty.
On the other hand, besides the alternating sign of the
norm of the states in the LD theory, the scheme may also
accommodate tachionic and/or massless states. In fact both the HD
and the LD formulations depend only on the {\it differences} of
the squared masses involved. So they are invariant
under the shifting of all the squared masses by an arbitrary
real quantity. Therefore any (but only one) of them can be
brought to zero, the greater ones remaining positive and the
lesser ones becoming negative (i.e. tachionic).

A key technique for the 2N-derivative theories, with $N\geq 3$,
is the use of Legendre
transformations involving analytical functions of the space-time
derivatives. The typical example is the definition of the
conjugate variable appearing in equation (3.3), namely
$\pi\equiv{{\partial{\cal L}_\psi}\over{\partial(\kg 1\gk\psi)}}
        = {{\partial{\cal L}^6_{\phi^o}}\over
{\partial(\kg 1\gk\kg 3\gk^{1\over 2}\phi^o)}}$. The mathematics
of this kind of transformations deserves further study in
relation with the formalism developed in refs.[10]. The
model presented here provides a working example.

An unavoidable feature of HD field theories is the occurrence of
negative norm (poltergeist) states, which is synonymous of
instability. The ensuing loss of unitarity seems hopeless
unless the full quantum corrections are
taken into account. Renormalization group calculations for
4-derivative gravity have failed to solve this difficulty.
The problem is intrinsically associated
to the finite differential order of the theory, but may be absent
if infinitely higher order terms are considered [11]
($N\rightarrow\infty$), as it is the
actual case of the effective theory stemming from the string and
quantum field theory in curved background.
Our simple scalar HD field model could provide a suitable test bed to
implement these ideas.

The occurrence of spurious fields is an unexpected byproduct and
is likely an artifact of our method. They are physically irrelevant once they
turn out to be decoupled from the source and from the other true
dynamical degrees of freedom. Several arguments stressing their
irrelevance have been presented above, stressing the idea the
they are indeed an artifact. A refined version of the
procedure we have followed might cope with them from the very
beginning at the price of losing some clearness of presentation.
It might also happen that they are naturally
absent in the framework of the alternative formulation of
Dirac's method for constrained systems as adopted in [12].
\bigskip\bigskip

\noindent{\bf Acknowledgements}

We are grateful to M.Le\'on for pointing out some relevant
references to us.

\vfill
\eject

\noindent{\bf Appendix A }

For use in the following we repeat the derivation of the
eqn's of motion from (2.14) back to (2.12), starting now from
the suitably modified 2--derivative field eqn's

$$\eqalign{ \kg 1\gk\phi^1_1 &= lj \cr
            -\kg 2\gk\phi^1_2 &= kj \quad,\cr}$$\hfill (A.1)

\noindent{where} $l$ and $k$ are arbitrary constants. In terms
of $\pi$ and $\phi^o$ they translate into

$$\eqalign{\kg 1\gk\pi &= -{1\over 2}\langle 12 \rangle
   \bigl(-\pi + {1\over 2}\phi^o\bigl) + {{l+k}\over 2}\,j\cr
           \kg 1\gk\phi^o &= \langle 12 \rangle
  \bigl(-\pi + {1\over 2}\phi^o\bigl) + (l-k)\,j\quad.\cr}$$\hfill (A.2)

Again, as in (2.13), working $\pi$ out of the 2nd equation (A.2) and
substituting it into
the 1st, one obtains the HD eqn.

$$ -{1\over \langle 12 \rangle}[\![1]\!][\![2]\!]\phi^o =
   \bigl(l - {{l-k}\over \langle 12
\rangle}[\![1]\!]\bigl)\,j \quad,$$\hfill (A.3)

\noindent{to} which corresponds the Lagrangian

$$ {\cal L}^{(4)}_{lk} = {1\over 2}{1\over \langle 12 \rangle}
                                \phi^o[\![1]\!][\![2]\!]\phi^o -
 \bigl(l - {{l-k}\over \langle 12
\rangle}[\![1]\!]\bigl)\,j \quad.$$\hfill (A.4)

\bigskip\bigskip

Let us come back now to the equations of motion of the
6-derivative theory with, eventually, a non-zero coupling constant
$\lambda$ of the spurious field to the source.

Equation (3.10) gets a term $-j\,\lambda \zeta^2_3$ so that
(3.9) has the new contribution $-j(\lambda C_2
{{\la 13\ra}\over{\la 23\ra}}\Phi^2_3+\lambda C_2\Theta^2_3)$.
Thus the equations of motion for $\Phi^2_3$ and $\Theta^2_3$
become

$$\eqalign {{1\over{\la 23\ra}}\kg 3\gk \Phi^2_3\,&=\,\Bigl(1+\lambda
C_2{{\la 13\ra}\over{\la 23\ra}}\Bigl)\,j\cr
-{1\over{\la 13\ra}}\kg 3\gk\Theta^2_3\,&=\,(1+\lambda
                                   C_2)\,j\quad, \cr}$$\hfill (A.5)

\bigskip

\noindent{which} together with the field equations for $\phi^2_1$
and $\phi^2_2$ derived from (3.9), namely

$$\eqalign {-{1\over{\la 23\ra}}\kg 1\gk\phi^2_1\,&=\,j\cr
             {1\over{\la 13\ra}}\kg
2\gk\phi^2_2\,&=\,j\quad,\cr}$$\hfill (A.6)

\bigskip

\noindent{and} equations (3.7),(3.8), combine to yield

$$\eqalign{{1\over{\la 13\ra\la 23\ra}}\kg 1\gk\kg 3\gk\phi^1_1\,&=\,
              \Bigl(1-\lambda{{C_2}\over{\la 23\ra}}\kg 1\gk\Bigl)\,j\cr
           -{1\over{\la 13\ra\la 23\ra}}\kg 2\gk\kg 3\gk\phi^1_2\,&=\,
             \Bigl(1-\lambda{{C_2}\over{\la 23\ra}}\kg
                               2\gk\Bigl)\,j\quad.\cr}$$\hfill (A.7)

\bigskip

\noindent{Using} (3.5) they may be written as

$$\eqalign {{1\over{\la 13\ra\la 23\ra}}\kg 1\gk\psi^1_1\,&=\,
                             l\; {\kg 3\gk}^{-{1\over2}}\,j\cr
          -{1\over{\la 13\ra\la 23\ra}}\kg 2\gk\psi^1_2\,&=\,
                              k\; {\kg
                         3\gk}^{-{1\over2}}\,j\quad,\cr}$$\hfill (A.8)

\noindent{where} $l=(1-\lambda{{C_2}\over{\la 23\ra}}\kg 1\gk)$ and
$k=(1-\lambda{{C_2}\over{\la 23\ra}}\kg 2\gk)$.
Now the same derivation that leads from (A.1) to (A.3) brings (A.8) to
the equivalent equation

$$ -{1\over{\la 12\ra\la 13\ra\la 23\ra}} \kg 1\gk\kg 2\gk \psi\,=\,
    \Bigl(l-{{l-k}\over{\la 12\ra}}\kg 1\gk\Bigl){\kg
                                 3\gk}^{-{1\over2}}\, j\quad,
                                                  $$\hfill (A.9)

\noindent{or}, with $\psi = {\kg 3\gk}^{1\over 2}\phi^o$, to the
final form

$$  -{1\over{\la 12\ra\la 13\ra\la 23\ra}}
                  \kg 1\gk\kg 2\gk\kg 3\gk\phi^o\,=\,
       \Bigl(l-{{l-k}\over{\la 12\ra}}\kg 1\gk\Bigl)\,j\quad .
                                                  $$\hfill (A.10)
\vfill
\eject

\noindent{But} substituting the actual values of $k$ and $l$ one
has that

$$  \Bigl(l-{{l-k}\over{\la 12\ra}}\kg 1\gk\Bigl)\,=\,1 \quad,
                                                 $$\hfill (A.11)

\noindent{so} that (A.10) is exactly the same unaltered HD field
equation (3.13).

\bigskip\bigskip

\noindent{\bf Appendix B}

To prove (4.8), some arrangements are in order.
First, the upper limit of the r.h.s. of (4.8) can be
extended to the pair $(N\; N+1)$ with the restriction $k,l \not= N+1$; this
property has already been used when getting (4.7) from (4.6).
Secondly, a factor $(-1)^{N+2}$ can be deleted and, after bringing all the
terms to the l.h.s., (4.8) adopts the simpler form

$$ \sum_{i=1}^{N+1}\;(-1)^{i-1}
  {\prod\limits_{(kl)=(12)\atop k,l\not=i}^{{ }\atop (N\;N+1)}\;
                                            \la kl\ra}  = 0
                                                 \quad.$$\hfill (B.1)

\noindent{This} compact version will be given a meaning in Appendix C.

Proving (B.1) can be better done by recasting it in an even more
convenient form. A common factor can be extracted from the sum,
obtaining

$$ {\prod\limits_{(kl)=(12)}^{{ }\atop (N\;N+1)}\; \la kl\ra}
     \Biggl[ \sum_{i=1}^{N+1}
    \Biggl({\prod\limits_{j=1\atop j\not=i}^{{ }\atop N+1}}
      {1\over{\la ij\ra}} \Biggl ) \Biggl ] = 0 \quad .$$\hfill (B.2)

\noindent{Notice} that in writing (B.2), the couples $\la ij\ra$
have been let not to respect the ordering convention $i<j$.
Thus, $i-1$ of them are negative, which explains why the sign
factor $(-1)^{i-1}$ in (B.1) does not occur in (B.2).

A sufficient condition for (B.2) to hold is that

$$  \sum_{i=1}^{N+1}
    \Biggl({\prod\limits_{j=1\atop j\not=i}^{{ }\atop N+1}}
      {1\over{\la ij\ra}} \Biggl )  = 0 \quad,$$\hfill (B.3)

\noindent{and} this will be proven again by induction. Let us
assume (B.3) to be valid for $N$ terms, namely

$$  \sum_{i=1}^{N}
    \Biggl({\prod\limits_{j=0\atop j+1\not=i}^{{ }\atop N-1}}
                      {1\over{\la i\;j+1\ra}} \Biggl )  =  0
                                           \quad,$$\hfill (B.4)

\noindent{in which} we have renamed $j$ by $j+1$.
{}From eq.(B.4), assigning to the indices $i=1,2,3,...,N$ new values
$i=1,3,4,...,N+1$ we also have

$$  \sum_{i=1\atop i\not= 2}^{{ }\atop N+1}
    \Biggl({\prod\limits_{{j=0\atop j+1\not= i}\atop j\not=1}^{{ }\atop N}}
                      {1\over{\la i\;j+1\ra}} \Biggl )  = 0
                                                   \quad,$$\hfill (B.5)

and with the values i=2,3,4,...,N+1 one obtains

$$  \sum_{i=2}^{{ }\atop N+1}
    \Biggl({\prod\limits_{j=1\atop j+1\not=i}^{{ }\atop N}}
                      {1\over{\la i\;j+1\ra}} \Biggl )  = 0
                                                  \quad.$$\hfill (B.6)

The equation (B.3) can be written as

$${1\over{\la 12\ra}}\Biggl[{\prod\limits_{j=3}^{{ }\atop N+1}}
                                           {1\over{\la 1j\ra}}
                           -{\prod\limits_{j=3}^{{ }\atop N+1}}
                                    {1\over{\la 2j\ra}} \Biggl]
 + \sum_{i=3}^{{ }\atop N+1}
    \Biggl({\prod\limits_{j=1\atop j\not=i}^{{ }\atop N+1}}
                              {1\over{\la ij\ra}} \Biggl) = 0\quad,
                                                   $$\hfill (B.7)

\noindent{or} as

$${1\over{\la 12\ra}}\Biggl[{\prod\limits_{j=2}^{{ }\atop N}}
                                       {1\over{\la 1\;j+1\ra}}
                           -{\prod\limits_{j=2}^{{ }\atop N}}
                                 {1\over{\la 2\;j+1\ra}}\Biggl]
 + \sum_{i=3}^{{ }\atop N+1}
    \Biggl({\prod\limits_{j=1\atop j\not=i}^{{ }\atop N+1}}
                              {1\over{\la ij\ra}} \Biggl) =0
                                  \quad.$$\hfill (B.8)

The two terms inside the squared brackets in (B.8) have been
arranged to coincide with the first ones in the sums of (B.5) and
(B.6) respectively, so (B.8)  can be written as

$${1\over{\la 12\ra}}\Biggl[- \sum_{i=3}^{{ }\atop N+1}
    \Biggl({\prod\limits_{{j=0\atop j+1\not= i}\atop j\not= 1}^{{ }\atop N}}
                              {1\over{\la i\;j+1\ra}} \Biggl)
              + \sum_{i=3}^{{ }\atop N+1}
    \Biggl({\prod\limits_{j=1\atop j+1\not= i}^{{ }\atop N}}
                              {1\over{\la i\;j+1\ra}} \Biggl)\Biggl]
   + \sum_{i=3}^{{ }\atop N+1}
    \Biggl({\prod\limits_{j=1\atop j\not= i}^{{ }\atop N}}
                              {1\over{\la ij\ra}} \Biggl) =0
                                                    \quad,$$\hfill (B.9)
\vfill
\eject

\noindent{and} renaming $j$ by $j+1$ in the last sum we get

$${1\over{\la 12\ra}}\Biggl[- \sum_{i=3}^{{ }\atop N+1}
    \Biggl({\prod\limits_{{j=0\atop j+1\not= i}\atop j\not= 1}^{{ }\atop N}}
                              {1\over{\la i\;j+1\ra}} \Biggl)
              + \sum_{i=3}^{{ }\atop N+1}
    \Biggl({\prod\limits_{j=1\atop j+1\not= i}^{{ }\atop N}}
                              {1\over{\la i\;j+1\ra}} \Biggl)\Biggl]
   + \sum_{i=3}^{{ }\atop N+1}
    \Biggl({\prod\limits_{j=0\atop j+1\not= i}^{{ }\atop N}}
                              {1\over{\la i\;j+1\ra}} \Biggl) =0
                                                \quad.$$\hfill (B.10)

Now we can check that (B.10) is true because it is exactly verified
for each fixed index value i=3,...,N+1.

For the case $N=2$ , we have

$$ {1\over{\la 12\ra\la 13\ra}} + {1\over{\la 21\ra\la 23\ra}}
                               + {1\over{\la 31\ra\la 32\ra}} = 0
                                             \quad.$$\hfill (B.11)

This is inmediately seen because (B.11) reduces to the trivial identity
$\la 23\ra - \la 13\ra + \la 12\ra = 0$ , after multiplying by
$ \la 12\ra\la 13\ra\la 23\ra$.

For any $ m\geq 3$ , we have in the same way that

$$ {1\over{\la 12\ra\la 1m\ra}} + {1\over{\la 21\ra\la 2m\ra}}
                               + {1\over{\la m1\ra\la m2\ra}} = 0
                                             \quad.$$\hfill (B.12)

Next we consider the terms with fixed $i = m \geq 3$ in (B.10), which we
also claim to add to zero, namely

$$  - {1\over{\la 12\ra}}{\prod\limits_{{j=0\atop j+1\not=m}
        \atop j\not= 1}^{{ }\atop N}}{1\over{\la m\;j+1\ra}}
    +  {1\over{\la 12\ra}}{\prod\limits_{j=1\atop j+1\not=m}
        ^{{ }\atop N}}{1\over{\la m\;j+1\ra}}
    +  {\prod\limits_{j=1\atop j\not=m}
        ^{{ }\atop N+1}}{1\over{\la mj\ra}}  = 0 \quad,$$\hfill (B.13)

\noindent{because}, with $n=j+1$ in the second and first terms,
(B.13) can be written as

$$  - {1\over{\la 12\ra}}{1\over{\la m1\ra}}
       {\prod\limits_{n=3\atop n\not= m}^{{ }\atop N+1}}
                                    {1\over{\la mn\ra}}
    + {1\over{\la 12\ra}}{1\over{\la m2\ra}}
       {\prod\limits_{n=3\atop n\not= m}^{{ }\atop N+1}}
                                    {1\over{\la mn\ra}}
    + {1\over{\la m1\ra}}{1\over{\la m2\ra}}
       {\prod\limits_{n=3\atop n\not= m}^{{ }\atop N+1}}
                                    {1\over{\la mn\ra}}  =0
                                                 \quad,$$\hfill (B.14)

\noindent{or} else as

$$\Bigl[{1\over{\la 12\ra\la 1m\ra}} + {1\over{\la 21\ra\la 2m\ra}}
                               + {1\over{\la m1\ra\la m2\ra}}\Bigl]
               {\prod\limits_{n=3\atop n\not= m}^{{ }\atop N+1}}
                                    {1\over{\la mn\ra}}  = 0 \quad ,
                                                      $$\hfill (B.15)

\noindent{which} trivially holds because of (B.12). Then (B.10)
is true, and we have
proven (B.3) for $N+1$ terms, provided it holds for $N$ terms;
but (B.3) is true
for $N=2$ , which is nothing but equation (B.11), so (B.3) is
satisfied for any $N$.
\bigskip\bigskip

\noindent{\bf Appendix C}

Consider the cohomological space spanned by the simplices

$$\eqalign{
             0-simplices\quad &{ }P_i \cr
             1-simplices\quad &(P_iP_j)\cr
             2-simplices\quad &(P_iP_jP_k)\cr
             3-simplices\quad &(P_iP_jP_kP_l)\cr
             ......\quad\quad &\quad......\cr}$$

\noindent{corresponding} to points $i=1,2,...\;;$ ordered couples of points
$i<j\;;$ ordered triads $i<j<k\;;$ ordered tetrads
$i<j<k<l\;;$ etc., and endowed with the boundary
operator $\partial{ }$ \space :

$$\eqalign{
          \partial P_i &= 0 \cr
      \partial(P_iP_j) &= P_i-P_j \cr
   \partial(P_iP_jP_k) &= (P_jP_k)-(P_iP_k)+(P_iP_j)\cr
\partial(P_iP_jP_kP_l) &= (P_jP_kP_l)-(P_iP_kP_l)+(P_iP_jP_l)-(P_iP_jP_k)\cr
      ......\quad\quad &\quad\quad\quad......\cr}$$

\noindent{It} can be trivially checked that $\partial^2 = 0$.

Any n-chain may be given a wheight by assigning the following
wheights to the simplices:

$$\eqalign{
           P_i{}&\longrightarrow m_i^2 \cr
         (P_iP_j)&\longrightarrow \la ij\ra\equiv m_i^2-m_j^2\cr
      (P_iP_jP_k)&\longrightarrow \la ij\ra\la jk\ra\la ik\ra\cr
   (P_iP_jP_kP_l)&\longrightarrow \la ij\ra\la jk\ra\la ik\ra
                                  \la il\ra\la jl\ra\la kl\ra\cr
     ......\quad &\quad\quad\quad......\cr}$$

Then equation (B.1) can be read as the following statement :
For $n\geq 1$, the wheight of any closed $n$-chain is zero .

\noindent{The} lower (trivial) case of this statement is
$ \la 23\ra-\la 13\ra+\la 12\ra = \la 12\ra+\la 23\ra+\la 31\ra = 0$.

\bigskip\bigskip

\noindent{\bf Appendix D}

We will prove the equivalence of (4.9) and (4.10) again by
the induction method. First  note that
(4.10) for $N=3$ is just (3.10) where the coefficient in the
spurious Lagrangian has been brought down to just $1\over2$ by a
rescaling of the spurious field. Next take

$$\psi = \Bigl[ \kg 3\gk\kg 4\gk\cdot\cdot\cdot\kg N+1\gk\Bigl]^
           {1\over 2}\;\phi^o \quad,$$\hfill (D.1)

\noindent{and}, following a similar procedure to the one used
in the 6-derivative theory, define

$$ \pi\equiv{{\partial{\cal L}^{2N+1}}\over{\partial(\kg
                            1\gk\psi)}}\quad,$$\hfill (D.2)

\noindent{go} to the Helmholtz Lagrangian, and choose

$$\eqalign{\psi_1^1 &=\pi + {1\over 2}\psi\cr
            \psi_2^1 &= -\pi + {1\over 2}\psi \quad.\cr}$$\hfill (D.3)

\noindent{The} lagrangian for the $2(N+1)$ case then reads:

$${\cal L}^{2(N+1)}=
        -{1\over 2}{\prod\limits_{(ij)=(13)}^{{ }\atop (N\;N+1)}}
                     {1\over{\la ij\ra}}\;\Bigl[(-1)\psi_1^1\kg 1\gk
                       \psi_1^1 + \psi_2^1\kg 2\gk\psi_2^1\Bigl]
     - j\,\Bigl[ \kg 3\gk\kg 4\gk\cdot\cdot\cdot\kg N+1\gk\Bigl]^
                                {-{1\over 2}}(\psi_1^1 + \psi_2^1)
                                            \quad, $$\hfill (D.4)
\vfill
\eject

\noindent{where} the factor ${1\over{\la 12\ra}}$ has been used
to split the KG operators $\kg 1\gk$  and $\kg 2\gk$. With

$$\eqalign{
 \phi_1^1 & = \Bigl[ \kg 3\gk\kg 4\gk\cdot\cdot\cdot\kg N+1\gk\Bigl]^
           {-{1\over 2}}\;\psi_1^1\cr
 \phi_2^1 & = \Bigl[ \kg 3\gk\kg 4\gk\cdot\cdot\cdot\kg N+1\gk\Bigl]^
           {-{1\over 2}}\;\psi_2^1\quad,\cr}$$\hfill(D.5)

\noindent{we} have

$$\eqalign{
        {\cal L}^{2(N+1)}=
        -{1\over 2}{\prod\limits_{(ij)=(13)}^{{ }\atop (N\;N+1)}}
                     {1\over{\la ij\ra}}\;\Biggl[(-1)\phi_1^1
\Bigl(\prod\limits_{k=1\atop k\not= 2}^{{ }\atop N+1}\kg k\gk\Bigl)
                       \phi_1^1
&+ \phi_2^1
\Bigl(\prod\limits_{k=2\atop{ }}^{{ }\atop N+1}\kg k\gk\Bigl)
                       \phi_2^1\Biggl]\cr
                             & - j\,\phi_1^1 -
                               j\,\phi_2^1\quad,\cr}$$\hfill (D.6)

\noindent{that} can be written as

$$\eqalign{
        {\cal L}^{2(N+1)}=
       &\Biggl[(-1){\prod\limits_{k=3\atop{ }}^{{ }\atop N+1}}
                     {1\over{\la 2k\ra}}\Biggl]
 \Biggl[-{1\over 2}{\prod\limits_{(ij)=(13)\atop{i,j\not= 2 }}
^{{ }\atop (N\;N+1)}}
                     {1\over{\la ij\ra}}\;\phi_1^1
\Bigl(\prod\limits_{k=1\atop k\not= 2}^{{ }\atop N+1}\kg k\gk\Bigl)
                       \phi_1^1\Biggl] -j\phi_1^1\cr
&+\Biggl[{\prod\limits_{k=3\atop{ }}^{{ }\atop N+1}}
                     {1\over{\la 1k\ra}}\Biggl]
 \Biggl[-{1\over 2}{\prod\limits_{(ij)=(23)\atop{i,j\not= 1 }}
^{{ }\atop (N\;N+1)}}
                     {1\over{\la ij\ra}}\;\phi_2^1
\Bigl(\prod\limits_{k=2\atop k\not= 1}^{{ }\atop N+1}\kg k\gk\Bigl)
                  \phi_2^1\Biggl] -j\phi_2^1\quad.\cr}$$\hfill (D.7)

Now, observe that inside the first bracket we have the expression
for a 2N derivative theory, with $\phi_1^1$  in the place of
$\phi^o$,  and with the
KG operators $\kg 1\gk \kg 3\gk\kg 4\gk\cdot\cdot\cdot\kg
N\gk$ $\kg N+1\gk$. The factor that multiplies the
kinetic term, does not play any role as in the N=3 case. Inside the
second bracket we have also a 2N derivative theory with
$\phi_2^1$ in place of $\phi^o$ ,
and with the operators$\kg 2\gk \kg 3\gk\cdot\cdot\cdot\kg
N\gk\kg N+1\gk$. Then, with the assumption that
(4.10) is true for the 2N-derivative theory, we have

$$\eqalign{
{\cal L}^{2(N+1)}=
       \Biggl[(-1){\prod\limits_{k=3\atop{ }}^{{ }\atop N+1}}
                     {1\over{\la 2k\ra}}\Biggl]
&\Biggl[{1\over 2}(-1)^N{\prod\limits_{(mn)=(34)\atop{m,n\not= 1,2}}
^{{ }\atop (N\;N+1)}}{1\over{\la mn\ra}}\;\phi_1^N\kg 1\gk\phi_1^N\cr
&+ {1\over 2}\sum_{i=3\atop i_1=i}^{{ }\atop N+1}(-1)^{N-i}
  {\prod\limits_{{(mn)=(13)\atop m,n\not= i}\atop m,n\not= 2}
  ^{{ }\atop(N\;N+1)}} {1\over{\la mn\ra}}\;\phi_{i_1}^N\kg i\gk\phi_{i_1}^N\cr
&+ {1\over 2}\sum_{M=4\atop{ }}^{{ }\atop N+1}
            \sum_{l_1=1\atop{ }}^{{ }\atop(2^{M-3}-1)} (-1)^a
      \;\zeta_{Ml_1}^N\kg M\gk\zeta_{Ml_1}^N\Biggl]
- j\;\Bigl( \phi_1^N +
        \sum_{i_1=3\atop{ }}^{{ }\atop N+1}\phi_{i_1}^N\Bigl)\cr
+ \Biggl[{\prod\limits_{k=3\atop{ }}^{{ }\atop N+1}}
                     {1\over{\la 1k\ra}}\Biggl]
&\Biggl[{1\over 2}(-1)^N{\prod\limits_{(mn)=(34)\atop{m,n\not= 1,2}}
^{{ }\atop (N\;N+1)}}{1\over{\la mn\ra}}\;\phi_2^N\kg 2\gk\phi_2^N\cr
&+ {1\over 2}\sum_{i=3\atop i_2=i}^{{ }\atop N+1}(-1)^{N-i}
  {\prod\limits_{{(mn)=(23)\atop m,n\not= i}\atop m,n\not= 1}
  ^{{ }\atop(N\;N+1)}} {1\over{\la mn\ra}}\;\phi_{i_2}^N\kg i\gk\phi_{i_2}^N\cr
&+ {1\over 2}\sum_{M=4\atop{ }}^{{ }\atop N+1}
            \sum_{l_2=1\atop{ }}^{{ }\atop(2^{M-3}-1)} (-1)^b
      \;\zeta_{Ml_2}^N\kg M\gk\zeta_{Ml_2}^N\Biggl]
- j\;\Bigl( \phi_2^N +
        \sum_{i_2=3\atop{ }}^{{ }\atop N+1}\phi_{i_2}^N\Bigl)\quad,\cr}
                                                       $$\hfill (D.8)

\noindent{where} $ a = N+1-M+l_1-1$  and $ b = N+1-M+l_2-1$ .

	To get to (D.8), one needs to notice that the number of the
spurious fields associated to the operator $\kg M\gk$ depends on the place it
occupies in the set $\kg 1\gk \kg 3\gk\cdot\cdot\cdot\kg N+1\gk$
or in $\kg 2\gk \kg 3\gk\cdot\cdot\cdot\kg N+1\gk$, that for
$i\geq 3$ is $M-1$; the same is true for the signs of the kinetic terms.

	In (D.8), we see that there are two contributions to the
i-th KG Lagrangian for $i\geq 3$ . To disentangle this point we define

$$\eqalign{\phi_i^N&\equiv\phi_{i_1}^N + \phi_{i_2}^N\cr
  \phi_{i_s}^N&\equiv C_{i_1}\phi_{i_1}^N + C_{i_2}\phi_{i_2}^N\quad,\cr}
                                          $$\hfill (D.9)
\vfill
\eject

\noindent{i.e.}
$$\eqalign{
\phi_{i_1}^N &= {1\over{C_{i_2}-C_{i_1}}}(C_{i_2}\phi_i^N - \phi_{i_s}^N)\cr
 \phi_{i_2}^N &= {1\over{C_{i_2}-C_{i_1}}}(\phi_{i_s}^N
                             -C_{i_1}\phi_i^N)\quad,\cr}$$\hfill (D.10)

\noindent{for} all $i\geq 3$ . The KG Lagrangians  for any
$i , i_1 , i_2 \geq 3$ can be then written as

$$\eqalign{
(-1)^{N-i+1}{1\over 2}&{\prod\limits_{k=3\atop{ }}^{{ }\atop N+1}}
{1\over{\la 2k\ra}}
{\prod\limits_{{(mn)=(13)\atop m,n\not= i}\atop m,n\not= 1}
^{{ }\atop (N\;N+1)}}{1\over{\la mn\ra}}\;\phi_{i_1}^N\kg i\gk\phi_{i_1}^N\cr
+ (-1)^{N-i}{1\over 2}&{\prod\limits_{k=3\atop{ }}^{{ }\atop N+1}}
{1\over{\la 1k\ra}}
{\prod\limits_{{(mn)=(23)\atop m,n\not= i}\atop m,n\not= 1}
^{{ }\atop (N\;N+1)}}{1\over{\la mn\ra}}\;\phi_{i_2}^N\kg i\gk\phi_{i_2}^N\cr}
$$

$$ = (-1)^{N-i+1}{1\over 2}
{\prod\limits_{(mn)=(12)\atop m,n\not= i}^{{ }\atop (N\;N+1)}}
{1\over{\la mn\ra}}\;\Bigl[{{\la 12\ra}\over{\la 2i\ra}}
\phi_{i_1}^N\kg i\gk\phi_{i_1}^N - {{\la 12\ra}\over{\la 1i\ra}}
\phi_{i_2}^N\kg i\gk\phi_{i_2}^N\Bigl]
                                              \quad,$$\hfill (D.11)

\noindent{and} with (D.10) we get

$$\eqalign{
              (-1)^{N-i+1}{1\over 2}
{\prod\limits_{(mn)=(12)\atop m,n\not= i}^{{ }\atop (N\;N+1)}}
{1\over{\la mn\ra}}\;
\Bigl[\Bigl({{\la 12\ra}\over{\la 2i\ra}}
                       {{(C_{i_2})^2}\over{(C_{i_2}-C_{i_1})^2}}
           -{{\la 12\ra}\over{\la 1i\ra}}
                       {{(C_{i_1})^2}\over{(C_{i_2}-C_{i_1})^2}}\Bigl)
                                            \phi_i^N&\kg i\gk\phi_i^N\cr
+ \Bigl({{\la 12\ra}\over{\la 2i\ra}}
                       {1\over{(C_{i_2}-C_{i_1})^2}}
           -{{\la 12\ra}\over{\la 1i\ra}}
                       {1\over{(C_{i_2}-C_{i_1})^2}}\Bigl)
                            \phi_{i_s}^N&\kg i\gk\phi_{i_s}^N  \cr
- {{\la 12\ra}\over{\la 2i\ra}}
                       {{2C_{i_2}}\over{(C_{i_2}-C_{i_1})^2}}
                           \phi_i^N\kg i\gk\phi_{i_s}^N
            + {{\la 12\ra}\over{\la 1i\ra}}
                       {{2C_{i_1}}\over{(C_{i_2}-C_{i_1})^2}}
                           \phi_i^N&\kg i\gk\phi_{i_s}^N \Bigl]\quad.\cr}
                                                   $$\hfill (D.12)

\noindent{Choosing}

$$ C_{i_1}= {{\la 1i\ra}\over{\la 2i\ra}}C_{i_2} \quad ,
                                         $$\hfill (D.13)

\noindent{the} crossed terms in (D.12) desappear, and after
inserting the expression for $C_{i_1}$ we get

$$(-1)^{N+1-i+1}{1\over 2}
{\prod\limits_{(mn)=(12)\atop m,n\not= i}^{{ }\atop (N\;N+1)}}
{1\over{\la mn\ra}}\;\phi_i^N\kg i\gk\phi_i^N
+ (-1)^{N-i+1}{1\over 2}
  {\prod\limits_{(mn)=(12)\atop m,n\not= i}^{{ }\atop (N\;N+1)}}
  {1\over{\la mn\ra}}\;{{\la 2i\ra}\over{\la 1i\ra}}
                       {1\over{(C_{i_2})^2}}
                            \phi_{i_s}^N\kg i\gk\phi_{i_s}^N
                                                $$\hfill (D.14)

	So we observe, that the kinetic term for $i\geq 3$ , is exactly the
one we need to reach (4.10) for $N+1$. On the other side, it is trivial to
check that the kinetic terms in (D.8), for $\phi_1^N$  and $\phi_2^N$
, are the appropiated ones to fulfill (4.10) for the desired $N$.

	A brief statistics of the spurious fields is in order.
In (D.14) we get a rather complicated coefficient that is
immaterial because we can arbitrarily normalize these fields
since they do not couple to the other fields and sources.
Just notice that it is positive for $i = N+1$ ,
and alternating in sign as $i$ gets lesser. The mass degeneracy
is the following: With mass $m_M$, i.e. associated to the KG
operator $\kg M\gk$ in (D.8), we have $ 2^{M-3}-1$ spurious
terms with positive norm and  $ 2^{M-3}-1$
with negative norm. Equation (D.14) yields a further one,
rising the total number to
$2^{M-2}-1$. For $i = N+1$ the positive terms outnumber the
negative ones by one unit,
with this balance alternating for dwindling $i$.

This proves (4.10), because it holds for $N+1$ if it does
for $N$, and the procedure to prove the case $N = 3$ is legitimate.
\bigskip\bigskip

\vfill
\eject

\centerline{REFERENCES}\vskip1.0cm

\noindent [1] D.J.Gross and E.Witten,{\it Nucl.Phys.}{\bf B277}(1986)1.

          R.R.Metsaev and A.A.Tseytlin,{\it Phys.Lett.}{\bf B185}(1987)52 .

           M.C.Bento and O.Bertolami,{\it Phys.Lett.}{\bf B228}(1989)348.

\noindent [2] N.D.Birrell and P.C.W.Davies,{\it Quantum Fields in
                 Curved Space},

                 Cambridge Univ.Press(1982).

\noindent [3] K.S.Stelle,{\it Phys.Rev.}{\bf D16}(1977)953.

\noindent [4] T.Goldman, J.P'rez-Mercader, F.Cooper and M.M.Nieto,
               {\it Phys.Lett.}{\bf 281}(1992)219.

\noindent [5] M.Ferraris and J.Kijowski,{\it Gen.Rel.Grav.}{\bf 14}(1982)165.

              A.Jakubiec and J.Kijowski,{\it Phys.Rev.}{\bf D37}(1988)1406.

              G.Magnano, M.Ferraris and M.Francaviglia,
                                  {\it Gen.Rel.Grav.}{\bf 19}(1987)465;

                                  {\it J.Math.Phys.}{\bf 31}(1990)378;
                                  {\it Class.Quantum.Grav.}{\bf 7}(1990)557.

\noindent [6] J.C.Alonso, F.Barbero, J.Julve and A.Tiemblo,
                              {\it Class.Quantum Grav.}{\bf 11}(1994)865.

\noindent [7] J.C.Alonso and J.Julve, {\it Particle contents of
                higher order gravity},

              in Classical and Quantum Gravity (proc. of the
              1st. Iberian Meeting on Gravity,

              Evora, Portugal,1992), World Sci.Pub.Co.(1993)301.

\noindent [8] N.H.Barth and S.M.Christensen,
                                  {\it Phys.Rev.}{\bf D28}(1983)1876.

\noindent [9] E.T.Whittaker, {\it A treatise on the analytical dynamics of
              particles and rigid bodies},

              Cambridge Univ. Press,(1904). Pioneeering work by
              M.Ostrogradski and W.F.Donkin

              is quoted here.

\noindent [10] D.J.Saunders and M.Crampin,
                                   {\it J.Phys}{\bf A23}(1990)3169.

               M.J.Gotay, {\it Mechanics, Analysis and Geometry:
                           200 Years after Lagrange},

                  M.Francaviglia Ed., Elsevier Sci.Pub.(1991).

\noindent [11] D.A.Eliezer and R.P.Woodard,
                             {\it Nucl.Phys.}{\bf B325}(1989)389.

               J.Z.Simon,{\it Phys.Rev.}{\bf D41}(1990)3720.

\noindent [12] S.Hamamoto, Toyama Univ. preprint (1995), hep-th/9503177.

\bye